\documentstyle[epsfig,amsbsy,a4,12pt]{article}

\begin{document}
\bibliographystyle{plain}

\newcommand{\Opal}{\mbox{O{\sc pal}}}
\newcommand{\LepII}{\mbox{LEP2}}
\newcommand{\LepI}{\mbox{LEP1}}
\newcommand{\Lep}{\mbox{LEP}}
\newcommand{\Jetset}{\mbox{J{\sc etset}}}
\newcommand{\Pythia}{\mbox{P{\sc ythia}}}
\newcommand{\Herwig}{\mbox{H{\sc erwig}}}
\newcommand{\Wopper}{\mbox{W{\sc opper}}}
\newcommand{\Com}{centre-of-mass}
\newcommand{\SM}{Standard Model}
\newcommand{\MC}{Monte Carlo}
\newcommand{\tgc}{{\small TGC}}
\newcommand{\Prob}{{\cal P}}
%
%
\newcommand{\GeV}{\mbox{$\mathrm{GeV}$}}
\newcommand{\GeVc}{\mbox{$\mathrm{GeV}\!/\!c$}}
\newcommand{\GeVcc}{\mbox{$\mathrm{GeV}\!/\!c^2$}}
\newcommand{\Ipb}{\mbox{pb$^{-1}$}}
%
%
\newcommand{\beq}{\begin{equation}}
\newcommand{\eeq}{\end{equation}}
\newcommand{\bea}{\begin{eqnarray}}
\newcommand{\eea}{\end{eqnarray}}
\newcommand{\ra}{\rightarrow}
\newcommand{\mbf}{\mathbf}
\newcommand{\bm}{\boldmath}
\renewcommand{\deg}{^\circ}
%
%
\newcommand{\etal}{{\it et al.}} 
\newcommand{\PLB}[3]  {Phys.\ Lett.\ {\bf B#1} (#2) #3}
\newcommand{\ZPC}[3]  {Zeit.\ Phys.\ {\bf C#1} (#2) #3}
\newcommand{\NIMA}[3] {Nucl.\ Instr.\ Meth.\ {\bf A#1} (#2) #3}
\newcommand{\PPE}[1]  {CERN-PPE/{#1}}
\newcommand{\PRLD}[3] {Phys.\ Rev.\ Lett.\ {\bf D#1} (#2) #3}
\newcommand{\PRL}[3]  {Phys.\ Rev.\ Lett.\ \textbf{#1} (#2) #3}
\newcommand{\PRD}[3]  {Phys.\ Rev.\ {\bf D#1} (#2) #3}
\newcommand{\NPB}[3]  {Nucl.\ Phys.\ {\bf B#1} (#2) #3}
\newcommand{\NPhys}   {Nucl.~Phys}
\newcommand{\CPC}[3]  {Comp.\ Phys.\ Comm.\ {\bf #1} (#2) #3}
%
%
%
\newcommand{\Mz}{\mbox{$M_{\mathrm{Z}^0}$}}
\newcommand{\Gz}{\mbox{$\Gamma_{\mathrm{Z}}$}}
\newcommand{\Mw}{\mbox{$M_{\mathrm{W}}$}}
\newcommand{\Gw}{\mbox{$\Gamma_{\mathrm{W}}$}}
\newcommand{\sw}{\mbox{$\sin\theta_w$}}
\newcommand{\cw}{\mbox{$\cos\theta_w$}}
\newcommand{\tw}{\mbox{$\tan\theta_w$}}
\newcommand{\swsq}{\mbox{$\sin^2\theta_w$}}
\newcommand{\cwsq}{\mbox{$\cos^2\theta_w$}}
\newcommand{\twsq}{\mbox{$\tan^2\theta_w$}}
\newcommand{\epem}{\mbox{$\mathrm{e^+e^-}$}}
\newcommand{\mpmm}{\mbox{$\mu^+\mu^-$}}
\newcommand{\tptm}{\mbox{$\tau^+\tau^-$}}
\newcommand{\lplm}{\mbox{$\ell^+\ell^-$}}
\newcommand{\Zz}{\mbox{${\mathrm{Z}^0}$}}
\newcommand{\WW}{\mbox{$\mathrm{W^+W^-}$}}
\newcommand{\Wm}{\mbox{$\mathrm{W^-}$}}
\newcommand{\Wp}{\mbox{$\mathrm{W^+}$}}
\newcommand{\Wpm}{\mbox{$\mathrm{W^\pm}$}}
\newcommand{\eeWW}{\mbox{\epem$\rightarrow$\WW}}
\newcommand{\eeffff}{\mbox{\epem$\rightarrow$4-fermions}}
\newcommand{\qq}{\mbox{$\mathrm{q\overline{q}}$}}
\newcommand{\ff}{\mbox{$\mathrm{f\overline{f}}$}}
\newcommand{\lnu}{\mbox{$\ell\overline{\nu}_{\ell}$}}
\newcommand{\lnubar}{\mbox{$\overline{\ell}\nu_{\ell}$}}
\newcommand{\lmnu}{\mbox{$\ell^-\overline{\nu}_{\ell}$}}
\newcommand{\lpnu}{\mbox{${\ell^{\prime}}^+ \nu_{\ell^{\prime}}$}}
\newcommand{\enu}{\mbox{$\mathrm{e\overline{\nu}_{e}}$}}
\newcommand{\mnu}{\mbox{$\mu\overline{\nu}_{\mu}$}}
\newcommand{\tnu}{\mbox{$\tau\overline{\nu}_{\tau}$}}
\newcommand{\WWqqqq}{\mbox{\WW$\rightarrow$\qq\qq}}
\newcommand{\WWqqln}{\mbox{\WW$\rightarrow$\qq\lnu}}
\newcommand{\WWqqen}{\mbox{\WW$\rightarrow$\qq\enu}}
\newcommand{\WWqqmn}{\mbox{\WW$\rightarrow$\qq\mnu}}
\newcommand{\WWqqtn}{\mbox{\WW$\rightarrow$\qq\tnu}}
\newcommand{\WWlnln}{\mbox{\WW$\rightarrow$\lnu\lnubar}}
\newcommand{\Wenu}{\mbox{$\epem \rightarrow \mathrm{W}\enu$}}
\newcommand{\ZZqqqq}{\mbox{$(\Zz/\gamma)(\Zz/\gamma)\rightarrow\qq\qq$}}
\newcommand{\ZZggqqqq}{\mbox{$(\gamma^*\Zz)(\gamma^*\Zz)\rightarrow\qq\qq$}}
\newcommand{\Zeeqq}{\mbox{$\Zz\epem\rightarrow\qq \epem$}}
\newcommand{\Zee}{\mbox{$\epem\rightarrow\Zz\epem$}}
\newcommand{\ZZ}{\mbox{$\epem\rightarrow\Zz\Zz$}}
\newcommand{\Zqq}{\mbox{$\Zz\rightarrow\qq$}}
\newcommand{\Zqqqq}{\mbox{$\Zz\rightarrow\qq\qq$}}
\newcommand{\Gqq}{\mbox{$\gamma\rightarrow\qq$}}
\newcommand{\ZGqq}{\mbox{$\Zz/\gamma\rightarrow\qq$}}
\newcommand{\Mjet}{\mbox{$M_{\mathrm{jet}}$}}
\newcommand{\MGCE}{\mbox{$M_{\mathrm{GCE}}$}}
\newcommand{\Ecm}{\mbox{$E_{\mathrm{cm}}$}}
\newcommand{\Ebeam}{\mbox{$E_{\mathrm{beam}}$}}
\newcommand{\half}{\mbox{$\textstyle\frac{1}{2}$}}
\newcommand{\boldp}{\mbox{\boldmath$p$}}
\newcommand{\costh}{\mbox{$\cos\theta$}}
\newcommand{\Nctrk}{\mbox{$N_{\mathrm{tracks}}$}}
\newcommand{\Necal}{\mbox{$N_{\mathrm{ecal}}$}}
\newcommand{\dedx}{\mbox{$\mathrm{d}E/\mathrm{d}x$}}
\newcommand{\Econe}{\mbox{$E_{\mathrm{cone}}$}}
\newcommand{\Esub}{\mbox{$E_{\mathrm{sub}}$}}
\newcommand{\Econep}{\mbox{$E_{\mathrm{cone}}/p$}}
\newcommand{\Esubp}{\mbox{$E_{\mathrm{sub}}/p$}}
\newcommand{\Eovp}{\mbox{$E/p$}}
\newcommand{\Nblk}{\mbox{$N_{\mathrm{blk}}$}}
\newcommand{\Ndx}{\mbox{$N_{\mathrm{d}E/\mathrm{d}x}$}}
\newcommand{\Nsdx}{\mbox{$N^{\sigma}_{\mathrm{d}E/\mathrm{d}x}$}}
\newcommand{\Evis}{\mbox{$E_{\mathrm{vis}}$}}
\newcommand{\Rvis}{\mbox{$R_{\mathrm{vis}}$}}
\newcommand{\pmis}{\mbox{$\vec{p}_{\mathrm{mis}}$}}
\newcommand{\pvis}{\mbox{$\vec{p}_{\mathrm{vis}}$}}
\newcommand{\Rmis}{\mbox{$R_{\mathrm{mis}}$}}
\newcommand{\Rmax}{\mbox{$R_{\mathrm{max}}$}}
\newcommand{\Yc}[1]{\mbox{$y_{\mathrm{#1}}$}}
\newcommand{\Ephot}{\mbox{$E_{\mathrm{phot}}$}}
\newcommand{\Egam}{\mbox{$E_{\gamma}$}}
\newcommand{\roots}{\mbox{$\sqrt{s}$}}
\newcommand{\rootsprime}{\mbox{$\sqrt{s^\prime}$}}
\newcommand{\Ctmis}{\mbox{$|\cos\theta_{\mathrm{mis}}|$}}
\newcommand{\Ptsum}{\mbox{$\sum p_T$}}
\newcommand{\Zgamma}{\mbox{$\Zz/\gamma$}}
\newcommand{\ACOP}{\mbox{$\phi_{\mathrm{acop}}$}}
\newcommand{\XT}{\mbox{$x_T$}}
\newcommand{\XONE}{\mbox{$x_1$}}
\newcommand{\XTWO}{\mbox{$x_2$}}
\newcommand{\MLL}{\mbox{$m_{\ell\ell}$}}
\newcommand{\MRECOIL}{\mbox{$m_{\mathrm{recoil}}$}}
\newcommand {\ee}         {\mathrm{e}^+ \mathrm{e}^-}
\newcommand {\emu}         {\mathrm{e}^{\pm} \mu^{\mp}}
\newcommand {\et}         {\mathrm{e}^{\pm} \tau^{\mp}}
\newcommand {\mt}         {\mu^{\pm} \tau^{\mp}}
\newcommand {\mm}       {\mu^+ \mu^-}
\newcommand {\tautau}     {\tau^+ \tau^-}
\newcommand {\lemu}       {\ell=\mathrm{e},\mu}
\newcommand{\kt}{$k_{\perp}$}
\newcommand{\Probe}{\mbox{$P({\mathrm{e}})$}}
\newcommand{\Probm}{\mbox{$P({\mu})$}}
\newcommand{\Wmv}{\mbox{$\mathrm{W}\rightarrow\mnu$}}
\newcommand{\Wev}{\mbox{$\mathrm{W}\rightarrow\enu$}}
\newcommand{\Wtv}{\mbox{$\mathrm{W}\rightarrow\tnu$}}
\newcommand{\Wten}{\mbox{$\mathrm{W}\rightarrow\tnu\rightarrow
(\enu\ataunu)\taunu$}}
\newcommand{\Wtmn}{\mbox{$\mathrm{W}\rightarrow\tnu\rightarrow
(\mnu\ataunu)\taunu$}}
\newcommand{\Wtelect}{\mbox{$\mathrm{W}\rightarrow\tnu\rightarrow
(\enu\ataunu)\taunu$}}
\newcommand{\WWtelect}{\mbox{$\WW\rightarrow\qq\tnu\rightarrow
\qq(\enu\ataunu)\taunu$}}
\newcommand{\WWtonep}{\mbox{$\WW\rightarrow\qq\tnu\rightarrow
\qq(\pi^\pm n\pi^0\ataunu)\taunu$}}
\newcommand{\Wtonep}{\mbox{$\mathrm{W}\rightarrow\tnu\rightarrow
(\pi^\pm n\pi^0\ataunu)\taunu$}}
\newcommand{\Wtthreep}{\mbox{$\mathrm{W}\rightarrow\tnu\rightarrow 
(2\pi^\pm\pi^\mp\ataunu)\taunu$}}
%
%
\newcommand{\pw}{\mbox{$p_{\mathrm{W}}$}}
\newcommand{\Cth}{\mbox{$\cos\theta$}}
\newcommand{\Cthw}{\mbox{$\cos\theta_{\mathrm{W}} $}}
\newcommand{\Cthst}{\mbox{$\cos \theta^{*} $}}
\newcommand{\Phist}{\mbox{$\phi^{*} $}}
\newcommand{\Cthstl}{\mbox{$\cos \theta_\ell^{*}$}}
\newcommand{\Phistl}{\mbox{$\phi_\ell^{*}$}}
\newcommand{\Cthstj}
{\mbox{ $\cos\theta_{\scriptscriptstyle \mathrm{jet}}^{*}$}}
\newcommand{\Phistj}{\mbox{ $\phi_{\scriptscriptstyle \mathrm{jet}}^{*}$}}
\newcommand{\Cthstjf}
{\cos \theta_{\scriptscriptstyle \mathrm{jet}}^{*~{\scriptscriptstyle fol} } }
\newcommand{\Phistjf}
{\phi_{\scriptscriptstyle \mathrm{jet}}^{*~{\scriptscriptstyle fol}}}
\newcommand{\OO}{{\cal O}}
\newcommand{\Sgtot}{\sigma_{tot}}
\newcommand{\Qjet}{Q_{\mathrm{jet}}}
\newcommand{\DQ}{\Delta Q}
\newcommand{\Lnln}{\lnu\lnubar}
\newcommand{\Qqln}{\qq\lnu}
\newcommand{\Qqen}{\qq\enu}
\newcommand{\Qqmn}{\qq\mnu}
\newcommand{\Qqtn}{\qq\tnu}
\newcommand{\Qqqq}{\qq\qq}
\newcommand{\Qqee}{\qq\ee}

\newcommand{\gz}{\mbox{$g_1^{\mathrm{z}}$}}
\newcommand{\kg}{\mbox{$\kappa_\gamma$}}
\newcommand{\kz}{\mbox{$\kappa_{\mathrm{z}}$}}
\renewcommand{\lg}{\mbox{$\lambda_\gamma$}}
\newcommand{\lz}{\mbox{$\lambda_{\mathrm{z}}$}}
\newcommand{\dgz}{\mbox{$\Delta g_1^{\mathrm{z}}$}}
\newcommand{\dk}{\mbox{$\Delta \kappa$}}
\newcommand{\dkg}{\mbox{$\Delta \kappa_\gamma$}}
\newcommand{\dkz}{\mbox{$\Delta \kappa_{\mathrm{z}}$}}
\newcommand{\dlg}{\mbox{$\Delta \lambda_\gamma$}}
\newcommand{\dlz}{\mbox{$\Delta \lambda_{\mathrm{z}}$}}

\newcommand{\xg}{\mbox{$x_\gamma$}}
\newcommand{\xz}{\mbox{$x_{\mathrm{z}}$}}
\newcommand{\yg}{\mbox{$y_\gamma$}}
\newcommand{\yz}{\mbox{$y_{\mathrm{z}}$}}
\newcommand{\zz}{\mbox{$z_{\mathrm{z}}$}}
\newcommand{\zop}{\mbox{$z'_1$}}
\newcommand{\ztp}{\mbox{$z'_2$}}
\newcommand{\zhp}{\mbox{$z'_3$}}
\newcommand{\dz}{\mbox{$\delta_{\mathrm{z}}$}}

\newcommand{\abf}{\mbox{$\alpha_{B\phi}$}}
\newcommand{\awf}{\mbox{$\alpha_{W\phi}$}}
\newcommand{\aw}{\mbox{$\alpha_W$}}
\newcommand{\modbf}{\mbox{$B\phi$}  }
\newcommand{\modwf}{\mbox{$W\phi$}  }
\newcommand{\modw}{\mbox{$W$}  }
\newcommand{\mln}{M_{\ell\nu}}
\newcommand{\mqq}{M_{\mathrm{q}\bar{\mathrm{q}}}}

\def\dwf{\tilde{\alpha}_{W\Phi}}
\def\dbf{\tilde{\alpha}_{B\Phi}}
\def\dbw{\tilde{\alpha}_{BW}}
\def\dw{\tilde{\alpha}_{W}}
\def\tk{\tilde{\kappa}}
\def\tl{\tilde{\lambda}}

\newcommand{\LL}{\log L}
\newcommand{\Pfull}{P_{\tiny full}}
\newcommand{\Ehtot}{E_{\tiny had}^{\tiny tot}}

\def\onehalf{{1\over 2}}
\def\vtau{\mbox{\boldmath $\tau$}}
\def\vW{\mbox{\boldmath $W$}}
\def\vvW{\tilde{\mbox{\boldmath $W$}}}
\newcommand{\nn}{\nonumber}

%
%
\begin{titlepage}
\vspace{50mm}
\bigskip
\bigskip
\begin{center}
 {\Large \bf Report of the working group  \\
    on the measurement of \\
  triple gauge boson couplings \\}
\end{center}
\bigskip
\bigskip
\begin{center}
F.~Berends,
T.~Bowcock,
D.~Charlton,
P.~Clarke$^*$,
J.~Conboy,
C.~Hartmann,
P.~Kyberd,
C.G~Papadopoulos$^*$,
H.T~Phillips,
R.~Sekulin,
A.~Skillman \\
\end{center}
{\it \small $^*$ convenors} \\
\bigskip
\bigskip

Abstract: The working group discussed
several aspects of triple gauge coupling analysis viewed in the light
of experiences with the first high energy data recorded at
energies above the W pair threshold. Some analysis methods were reviewed
briefly, and consideration given to better ways of characterising the data.
The measurement of CP violating parameters was discussed.
Results were prepared to further quantify the precision attainable on 
anaomalous couplings in the four-quark channel using jet-charge methods, 
and finally the trade off between maximum LEP energy-vs-luminosity
was quantified.

\end{titlepage}

\tableofcontents

\pagebreak

\section{Introduction}
\label{sec-introduction}

The triple-gauge-coupling (TGC) working group was composed of 
members of the theoretical and
experimental community who are actively involved with the measurement and
interpretation of \tgc s at \LepII. Many of the participants 
also contributed to the \LepII\ workshop~\cite{LEP2YR} in 1995.

A review of the current status of the first experimental results
from \LepII\ 
was given in the plenary review [R.Sekulin - these proceedings] which
also gives an overview of some of the calculational tools
presently available as well as methods of analysis which have
been used.

At the \LepII\ workshop many prophecies were made in advance of the
first data taking above the W pair threshold.
The group decided that the best use of this workshop would be to
review how some of the ideas envisaged at that time had actually
evolved in the light of experience with the first data, and 
to discuss some of the problems which have been brought to light.
We also identified some quantatitive studies
which were considered useful to present for future reference.

In section two we consider briefly some of the limitations of 
present analysis methods.  
We then consider in some detail how experimental measurements
might be better characterised in order to avoid,
or minimise, some of these limitations in an unbinned maximum likelihood
fit. Finally the possibilities for an improved fitting tool are described. 

In the third section we 
describe a development of a method employing optimal observables, which
makes use of fully simulated distributions to extract \tgc\ parameters.

In the fourth section a study is presented in which the precision
which can be obtained using the \Qqqq\ channel
is quantified in the light of current experience of jet
pairing and charge assignment purity.  

In section five we present a parametrization of the CP violating
trilinear gauge couplings restricting ourselves to the gauge invariant
operators with the lowest dimension. We also present some estimate
of the sensitivity expected at LEP2. 

Finally in section six we present a study to quantify the variation
of \tgc\ measurement precision which would be obtained 
for different scenarios of
maximum \LepII\ energy versus the luminosity increase which 
results from sacrificing a few GeV from the absolute maximum 
energy which \Lep\ can achieve.

Throughout the following we make use of abbreviations for different
W pair decay types.
These are: 
(i) $\Qqqq$ events where both W bosons decay to a quark-antiquark 
final state,
(ii) $\Qqln$  events where one W decays to a 
quark-antiquark final state, and the other W decays to an 
electron, muon or tau, plus a neutrino and
(iii) $\Lnln$ events where both W bosons decay leptonically.
We also refer to two distinct ways in which 
the phase space variables are normally characterised
for each event.
The first is to use quantities which are very close to
those measured by the detectors, such as 
the four-vectors of the observed ``fermions'' (leptons
and jets), this is referred to as {\it 4-vector} analysis. 
The second is
to instead use the set of five angles given by the $\Wm$ flight
direction with respect to the incoming $\em$ beam direction, 
and the polar and azimuthal angles of the decay products of 
each W measured in the rest frame of the  respective W. 
A full description of these angles can be found in 
\cite{LEP2YR,BILENKY,SEKULIN}.
These angles are commonly written as 
\{$\Cthw,\Cthst_1,\Phist_1,\Cthst_2,\Phist_2$\} and methods using
these are referred to as {\it 5-angle} analyses.
We refer to both schemes generically as the set
\{$\Phi$\}.

\section{Data characterisation for maximum likelihood fitting. }
\label{sec-analtools}

Previous studies~\cite{LEP2YR} have examined possible observable
effects from anomalous \tgc s.
Anomalous \tgc s can affect both the total
production cross-section and the shape of the differential
cross-section as a function of the W$^-$ production angle.
Additionally, the relative contributions of each helicity state of 
the W bosons are changed, which in turn affects the distributions 
of their decay products.

The most straightforward approach which has been considered
is to use an unbinned multi-dimensional maximum likelihood method (ML)
where one identifies the most likely measurement of
a \tgc\, $\alpha$, by varying $\alpha$ to minimise the quantity
\[
-\LL = - \sum_{events~i} log P(\Phi_i,\alpha)
\]
where $\Phi_i$ represents the phase space measurements for
event $i$.
To do this is is necessary to have some function, $P$, which
returns the probability density function for a particular 
$\Phi_i$ to occur, and this can be conventionally derived as
the normalised differential cross section. 
It is reasonably straightforward to code an analytic
form for $P$ if one restricts to on-shell formula
without initial state radiation (ISR). 
The relevant expressions can be found in~\cite{BILENKY}
and an example code is given in ~\cite{KNEURCODE}. Such approaches
are simple, not very CPU intensive, but as well as neglecting
ISR and $\Gw$ also neglect detector acceptance and resolution. The neglect
of such effects was shown~\cite{LEP2YR}  to lead to biases of roughly the same
magnitude as the expected eventual experimental precision
for $500\Ipb$.
 
More sophisticated approaches have been implemented, which
attempt to allow for all of the above effects. A typical solution
is to compare an observed data distribution with that predicted
using fully simulated \MC\ events passed through the same 
selection requirements as the data. By using \MC\
distributions corresponding to different generated values 
of $\alpha$ one can make fits in order to measure the most likely
true value of $\alpha$. 
Examples of this are 
(i) a coarse binned maximum likelihood analysis~\cite{OPAL-TGC}
(ii) use of distributions of optimised observables~\cite{DELPHI-TGC},
(iii) $\chi^2$ fits to a reduced number of variables~\cite{ALEPH-TGC} and
(iii) nearest neighbour counting in \MC\ distributions~\cite{L3-TGC}.
These methods are straightforward in the case of one variable,
but tend to require either a very large number of \MC\ events as the
number of variables increases, or to necessitate very coarse binning
(the possible exceptions to this are the optimised observables methods
described below). There are various ways to try to 
make these methods more efficient,
however we decided to turn our attention back to the more
straightforward maximum likelihood approach and consider how
its failings could be rectified.

In order to perform a ``complete'' ML analysis one would like to include
all of the effects of ISR, $\Gw$, detector resolution and detector 
acceptance into a likelihood function which we will call $\Pfull$. 
The issues involved
in doing this fall into three classes: (i) to include ISR requires
the inclusion of some internal integration over a radiated photon
spectrum (ii) to include the effect of $\Gw$ depends upon whether
one attempts to supply W mass information with each data event 
or instead wishes to internally integrate over the W virtualities and (iii)
the facilities which can be included in $\Pfull$ depend upon  
the choice of quantities which one measures for each data event. 

To completely specify an event requires 8 pieces of information. 
Each of the four decay fermions requires 3 quantities (masses
are fixed). The total energy and momentum provide 4 constraints
(assuming that the 4-momentum of any ISR photon is known). 
The total number of pieces of information required is
therefore $4\times3 -4  = 8$ 
(although the overall azimuthal angle of the event
carries no information). 
In the 5-angle formulation the missing information is
equivalent to ignoring the two W masses. 
It is also necessary to know the total
energy and momentum to calculate the decay angles, 
and it is normally assumed that there is no ISR. 
This makes it impossible to then carry out a consistent integration
over ISR in $\Pfull$.
Conversely if the 4-vectors of all of the decay fermions are specified
then these completely determine both the W masses and 
the momentum carried by ISR. There is therefore no flexibility
left for these to be integrated over in $\Pfull$.

Another important consideration is that
if one chooses to characterise each event into 5-angles then
all the quantities are complicated transforms of the basic detector
measurements such as lepton energy and direction etc.
As a result the 5-angles are highly correlated and rarely
have Gaussian resolutions. Both of these facts make it non-trivial 
to incorporate integrations over detector resolution
or acceptance. 
If instead one chooses to characterise each event using 
the 4-vectors then some of these limitations are removed. 

The group therefore considered the most desirable way 
to characterise each type of event from an experimental point of
view. This was done separately for each W pair decay channel,
and the results are summarised in table \ref{tab-summary}.

\subsection{$\mbf \Qqen$ and $\mbf \Qqmn$ events}

In these events the lepton is normally well measured and
can be characterised by its energy, $E_l$, and its polar and
azimuthal angles, $\Cth_l$ and $\phi_l$. 
These three quantities
are largely uncorrelated and often have resolutions which
may be approximated 
\footnote{
  The energy will either be measured in a calorimeter or by
  the magnetic tracking detector, in which case either
  the measurement or its inverse is approximately Gaussian. 
}
as being Gaussian.

The hadronically decaying W results in two jets. It is often 
(although not invariably) the case that the jet angle measurements are
better than energy resolutions. It is therefore highly likely that
any apparent jet mass is mainly determined by energy resolution rather than
any physical mass. It therefore also follows that the measured
invariant mass of the two jet system is poorly related to the
true W virtuality on an event by event basis. 
This last point is important, as several fitting tools which exist
take as input the pairs of measured 4-vectors belonging to each W, 
and then interpret the invariant mass of each pair literally
when calculating the matrix element for the decay. 
This means that the inclusion of the effect of $\Gw$ is not via
any integration internal to the fitting tool, but comes from the
actual mass distribution of the data being fitted, 
and this is determined almost entirely by resolution errors.
Although at this time no quantative study has been done, it seems 
unlikely that this is sensible. It is therefore 
desirable to have the option to allow $\Pfull$ some freedom
to integrate internally over W virtuality.
We therefore concluded that the two jets should be characterised by 
their directions, $\Cth_1, \phi_1, \Cth_2$ and $\phi_2$
and optionally the sum of both jet energies, $\Ehtot$.

This choice of these variables is designed to allow those that are well
measured to be used, and those that may not we well measured
to be integrated over easily. 
If $\Ehtot$ is used then 8 quantities are specified.
Since it is not required to assume the total energy and momentum of the event
to specify the 8 quantities, then the freedom is left for 
$\Pfull$ to either assume no
ISR or to integrate over a photon radiation spectrum.
However there is no freedom left to integrate over the W
virtualities.
If, however, it is considered that the hadronic energy measurement is 
relatively poor then one can either perform an explicit
integration over the assumed resolution (see section \ref{sec-incdeteff})
or allow a complete integration over all possible values of $\Ehtot$ to
be included in $\Pfull$.

\subsection{$\mbf \Qqtn$ events}

These events differ from the previous category because one observes
only the tau decay
\footnote{
 It was not considered sensible (at this stage at least) to consider
 incorporating an integration over all tau decay modes 
 into $\Pfull$. 
}
products. 
It is likely that the direction of the 
decay products can serve as a useful approximation to the
actual tau direction (with a larger resolution) but the energy
will always be low as there is always at least one missing neutrino.
Therefore the simplest way to specify the event is to use exactly
the same quantities as for the \Qqen\ and \Qqmn\ events, but in addition
allow the integration over $E_l$ in $\Pfull$.

\subsection{$\mbf \Lnln$ events}

Studies have indicated that in the case of
electron and muon events there is 
useful information on anomalous couplings.
It is in principle possible to fully reconstruct~\cite{LEP2YR}
these events up to a two fold ambiguity using the two observed leptons, 
the beam energy and momentum, and the W mass. 
However there are several problems in doing this due to 
experimental resolution and that one has no knowledge
of the actual W masses for each event.
The working group therefore considered that it would
be better to use only the observed
quantities and integrate over everything else in $\Pfull$.
Thus such events would be characterised by 
$E_{l^+},\Cth_{l^+},\phi_{l^+},E_{l^-},\Cth_{l^-}$ and $\phi_{l^-}$
(note again that only the relative $\phi$ carries information).
If in addition an integration over $E_{l^+}$ and $E_{l^+}$ can
be included then tau events could also be used.

\subsection{$\mbf \Qqqq$ events}

These are the most problematic. The experience of the working group
from various \MC\ studies is that the W production direction
can be well measured from these events, but as for the \Qqln\
events there is probably no useful information about the individual W
masses on an event by event basis. 
However, following the previous reasoning, we might think of specifying
all of the jet angles, $\Cth_i, \phi_i,~~~i=1,4$. 
In this case we would have given all 8
pieces of information, hence completely specifying the W masses,
once the total energy and momentum are fixed (in fact this is used by
DELPHI in one of their mass measurement methods).
At the present time no detailed studies have done to resolve this
point from the point of view of \tgc\ analysis, and further work
is required.

\subsection{Inclusion of detector resolution and acceptance}
\label{sec-incdeteff}

The quantities chosen to specify each type of event are largely
uncorrelated and are often described by Gaussian resolutions. 
As such it is possible to imagine performing
a simple Gaussian integration over each to allow 
for detector resolution. 
This can be done externally to $\Pfull$ using commonly available
integrating functions.

Exactly how many quantities to integrate over depends
upon the event type.  In the case of \Qqln\ events it may be 
that the precision of the lepton measurements is more than adequate
for no resolution on either energy or direction to be included.
In this case one needs to perform an integration
over only the jet angles.  By the same arguments it may be
that in the case of \Lnln\  events resolution can be neglected entirely.
Exactly how to proceed in each case is an experimental question,
and also depends upon the statistical precision of the data.

Similar approximations may be used in order to account for detector
acceptance. For instance detectors normally have a sharp cut off
in $\Cth$ for lepton identification, but otherwise are
near 100\% efficiency and uniform in $\phi$. A simple
approximation such as this can be included in the 
normalisation of $\Pfull$. In the case of jets the situation
is not always so simple, but it may nevertheless be appropriate
to apply a sharp fiducial cut such that jets are only
used in a region where they are well reconstructed with high efficiency.

The treatment of resolution and acceptance suggested here is
an experimentally motivated approximation, and as such will never be
``exact''. However since the biases incurred by completely
ignoring these effects are only expected to be approximately the size of
the final \LepII\ statistical error, it may be that the level of 
approximation
involved here is adequate for all practical purposes.
Further studies are needed to quantify this issue.

\subsection{Prospects for a new semi-analytic analysis tool}

A feasibility study of a semi-analytic analysis tool has been
made. For this purpose two different problems should be
dealt with. In the first place an efficient program should be
developed which for each set of variables from Table 1 
calculates the probabilities. In that evaluation the effects 
of the W-width and ISR should be taken into account. Secondly,
it should be studied whether this analyzer properly determines
the TGC parameters from a Monte Carlo sample of unweighted
events.

The first issue, developing a program which calculates cross 
sections differential in the appropriate variables has been
accomplished. In fact, two independent programs have been made
in order to have cross checks. One program uses the matrix
elements of ERATO, the other of EXCALIBUR. All kinematical
variables of table 1 have been implemented, except the case
with the tau-lepton. Without ISR the chosen variables require
two, one
integration, sometimes none. With ISR two more integrations
have to be carried out.

The second issue, whether maximum likelihood fits based on 
probabilities related to the above specific kinematic variables
can really determine TGC parameters is at present under study.
One generates unweighted events, uses of these events only
the variables of table 1 and calculates the likelihood
function for e.g. nine values of a specific TGC parameter.
From this the TGC parameter is determined. In this way 
one could try to determine any
parameter on which the matrix element depends. 

In fact, so far the most extensive studies have been made
for a parameter other than a TGC. The W-mass has
been chosen as first parameter to be studied. One reason
for this choice is that the question of experimentally
motivated quantities can also be relevant for the mass
determination. Another is, that one may eventually like
to determine simultaneously both the mass and a TGC coupling. 

So, as a first test of the method the mass determination
from experimentally motivated quantities was chosen. To
this end typically 1600 unweighted events were generated with
ERATO or EXCALIBUR. The semi-leptonic case has been studied
in great detail. The data samples are CC03, CC10 and CC20 based
sets including ISR and have been analyzed with the same
matrix elements or simpler ones (e.g. a data set with ISR analyzed
without ISR).

The following results can be reported. Within the errors,
typically 35-50 MeV, the input mass could be reconstructed
when the same matrix element is used in the analysis as in the
generation of the sample. When simpler matrix elements in the
analysis than in the generation are used, shifts can arise.
Large shifts (300 MeV in some cases) may occur due to the neglect
of ISR in the analysis. Smaller shifts (up to 50 MeV) are 
found when a CC20 sample is analyzed with CC03 probabilities.

The mass determination is also successful in the four quark
case when one assumes to know which quark pair comes from a 
W. This reduces the folding. 
When the method is applied to the purely leptonic case the
mass reconstruction is not efficient, as one would
expect as there is less kinematical information.

The results on the mass determination will be published in
more detail~\cite{NEWFITTER} elsewhere. It seems that the maximum likelihood
method to determine $M_W$ from experimental motivated 
quantities is a useful addition to the existing direct
reconstruction method.

As to the TGC determination, the first results look promising,
but more results should be obtained before a detailed
discussion of the merits of the method can be presented.  
However it already now is clear that the method takes ISR and
full matrix elements into account in a very satisfactory way.

\begin{table}[htbp]
 \begin{center}
 \begin{tabular}{||l|c|c||} \hline \hline
  Channel  & Well measured  & ``Poorly measured''      \\
           &                & (may be integrated over) \\ 
\hline \hline
\Qqen    &$ E_l,\Cth_l,\phi_l,\Cth_{1},\phi_{1},\Cth_{2},\phi_{2}  $&$ \Ehtot $\\
\hline
\Qqmn    &$         ''                                                  $&$  ''    $\\
\hline
\Qqtn    &$ \Cth_l,\phi_l,\Cth_{1},\phi_{1},\Cth_{2},\phi_{2}  $&$ E_l,\Ehtot $\\
\hline
\Lnln    &$ E_{l^+},\Cth_{l^+},\phi_{l^+},E_{l^-},\Cth_{l^-}, \phi_{l^-}  $&$   -  $\\
\hline
\Qqqq    &$  \Cth_i, \phi_i,~~~i=1,4                      $&$        -             $\\
\hline \hline
 \end{tabular}
 \end{center}
\caption{ Experimentally motivated quantities to specify
for different W pair event categories.
}
\label{tab-summary}
\end{table}

\section{Development of optimal observable analysis.}
\label{sec-newdev}

In section \ref{sec-analtools} several methods of \tgc\ analysis
were mentioned which employ fully simulated \MC\ events in order
to analyse the data.
These methods include all
effects of ISR, $\Gw$, detector acceptance and resolution.
One problem which often arises is the number of bins necessary
in a multidimensional analysis. However 
methods which use ``optimal observables''  
have the potential to avoid this problem.

The use of optimal observables (OO) has been discussed widely in
the literature~\cite{LEP2YR,DIEHL,PAPADOPOULOSOO,DIEHL2}. 
The method consists in constructing observables, namely functions of the
observed kinematic variables $\Phi$, which maximise the sensitivity
to anomalous couplings 
There is one observable per coupling, $\OO_\alpha(\Phi)$,
given by:

\begin{equation}
\label{eqntgc-oo-defn}
\OO_\alpha(\Phi) = \frac{\frac{d\sigma_1}{d\Phi}}{\frac{d\sigma_0}{d\Phi}} 
\end{equation}
where the functions $d\sigma_i \over d\Phi$ are obtained from the quadratic
form of the Born level cross section:
\begin{equation}
\label{eqntgc-oo-expan}
  \frac{d\sigma}{d\Phi} = \frac{d\sigma_0}{d\Phi} 
                       +  \alpha \frac{d\sigma_1}{d\Phi}
                       +  \alpha^2  \frac{d\sigma_2}{d\Phi} 
\end{equation}

The observable is calculated for each event from the measured 
$\Phi$ variables.
In principle the mean value of  $\OO$ (averaged over the data set)
contains all of the information which can be extracted from 
the $\Phi$ distribution of the data,
provided that it is small enough that the neglect of the second order term
has a negligible effect.
The spectrum of the optimal observables might be regarded as a
projection of the five-dimensional differential cross section in 
$\Phi$ onto a direction which optimises the sensitivity.

As originally envisaged and discussed in the literature, the mean
value is used to determine $\alpha$ through
the inversion of a matrix relation between the two.
The effects of ISR and $\Gw$ can only be included if the the
resulting functional form of the differential cross section of
equation \ref{eqntgc-oo-defn} is known, which is not generally the case,
and therefore biases are introduced.
It is also in principle possible to include detector effects in the
procedure, but this can lead to problems of instability.

Instead of following this procedure a recent OPAL 
analysis~\cite{OPAL-TGC} adopts
a different approach. 
A similar method has also been developed by DELPHI~\cite{DELPHIOO}.

\begin{figure}[htbp]
\epsfxsize \the\textwidth
\epsffile{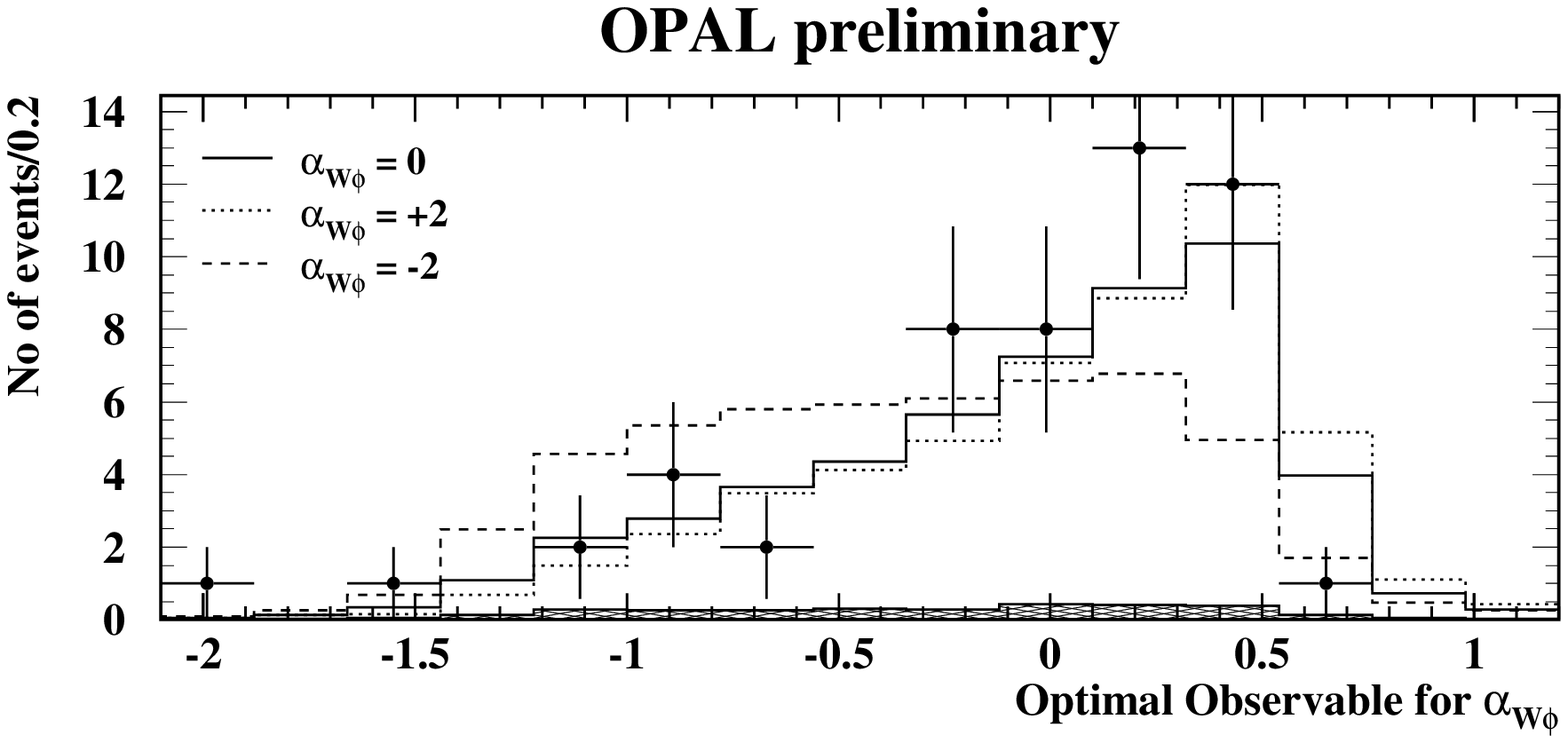}
\caption{ The distribution of the optimal observable for $\awf$
obtained from the $\Qqln$ events. 
The hatched histogram shows the non WW background.
These are compared with the distribution
expected in the \SM\ using fully simulated \MC\ events.
The predicted distributions for $\awf= +2 (-2)$ are also shown as
dotted (dashed) lines.}
\label{figtgc-ooplot}
\end{figure}

In the OPAL analysis 
each event is analysed using the 5-angles as the phase space variables.
Each OO quantity is calculated using the 
the on-shell/no ISR differential cross section
(it is straightforward to obtain an analytic function for this). 
The value of the OO measured per event is then 
treated just as any other experimental quantity whose distribution is
sensitive to variations in the relevant anomalous coupling. 
As example the observed distribution of the optimal observable for the 
$\alpha_{W\phi}$ parameter is shown in figure \ref{figtgc-ooplot} for data and \MC.
The expected distribution for all values of $\alpha$ is obtained using 
fully simulated EXCALIBUR~\cite{EXCALIBUR} events generated for a few specific
values, and then using the quadratic
dependence upon $\alpha$ in each bin of the spectrum of $\OO_\alpha$.
The \MC\ distributions are generated with ISR and $\Gw$ switched on.
The contribution from background contamination is 
added to the spectrum.

To measure the \tgc\ parameter a binned likelihood fit of the 
observed data to the  
expected distribution predicted by the \MC\ is performed. 
Since like is compared with like the method includes all of the 
required effects and no bias should be introduced.

There are two drawbacks of the method. The first is that
the OO is calculated using an on-shell/no ISR differential cross section.
This can be considered as an approximation to the true OO.
The approximation does not introduce any bias, nor should it be confused
with the fact that the data is fitted to \MC\ which includes ISR
and $\Gw$, however it does in principle lead to loss of optimality
which means that the measured statistical error will increase.
It is for this reason that the full distribution of the 
$\OO_\alpha$ calculated for each event is used, making better 
use of the available information than the mean value in the case
of this non-optimality. 
The second drawback is that the simple OO method is based upon an 
expansion about a single value
\footnote{ 
  Alternately, an extension of the method is possible~\cite{DIEHL2} 
  whereby the $\alpha^2$ term is incuded. 
}
of a given \tgc\ and is only optimal if the fitted values lie
in a small range
\footnote{
 If the measured central value, $\alpha_0$, 
 is different from the expansion point of $\alpha=0$
 then the method can be iterated to use OOs calculated from
 a functional form of equation \ref{eqntgc-oo-expan} where
 $\alpha$ is replaced by $\alpha - \alpha_0$.
}
around this value. At present the statistical precision
of the data is rather poor and so the OO method will not necessarily
give the true confidence levels for large values of the errors. 
However this limitation is expected to diminish as the statistical
precision of the data increases in future years.
However this method is in practice almost
adequate to analyse the current small data set. 
The $\LL$ distributions obtained
from the method described, and from a more conventional 
binned maximum likelihood analysis of the same data 
give almost identical errors and central values. These
$\LL$ curves can be seen in reference \cite{OPAL-TGC}.

\section{The use of jet charge in $\mbf \Qqqq$ events}

In events where both W pairs decay hadronically the simple method of determining
the W charge used in the semi-hadronic and leptonic events is no longer
available. A method of assigning a charge to the Ws from the properties of
the jet pairs into which they decay must be sought. Various possibilities 
suggest
themselves, but none of them lead to identification of the W charges with 100\%
certainty. Previous studies have either assumed that the charge is
unknown or completely determined; here we investigate the effect that varying
the efficiency of the W charge determination has on the ability to measure 
\awf. 
We generated events with \Pythia. The correct pairing of jets into Ws was
taken and, for a fraction $(1-f_g)$ of the events (chosen randomly), the correct
charge assignment of the Ws was taken. For the remaining fraction $f_g$ of the
events, the pair constituting the \Wm\ was called \Wp\ and vice-versa. This
procedure defines the ``assigned momenta'' with a probability $f_g$ of assumed
incorrect jet charge assignment.

In the analysis of the events, the unnormalized probability for observing each
event was taken as $(1-f_a) p(1,2) + f_a(2,1)$, where $p(i,j)$ is a squared
matrix element. For $(i,j) = (1,2)$ the production and decay angles of
\Wm\ and \Wp\ were calculated using the ``assigned momenta'' defined above,
while for $(i,j) = (2,1)$ the production and decay angles were taken with the
``assigned'' \Wm\ interpreted as the \Wp\ and vice-versa.

Results of extended maximum likelihood fits of \awf\ to 1000 events at 190 GeV
are shown in table~\ref{tab-misas}.
The decay angles from both Ws were included in the fits, but folded, thus
assuming no quark flavour information in the W decay products. Thus each 
$p(i,j)$ is an average of four matrix element calculations.

\begin{small}
\begin{table}[htbp]
 \begin{center}
 \begin{tabular}{||c||l|l|l|l|l|l||}
\hline \multicolumn{1}{||c||}{$f_a$, Rate} 
&\multicolumn{6}{c||}{$f_g$, Misassignment rate applied to generated events}\\ 
\cline{2-7}
 assumed &\multicolumn{1}{c|}{0.0} &\multicolumn{1}{c|}{0.1}
         &\multicolumn{1}{c|}{0.2} &\multicolumn{1}{c|}{0.3}
         &\multicolumn{1}{c|}{0.4} &\multicolumn{1}{c||}{0.5}                 \\
 in analysis & & & & & &                                              \\ \hline
  0.00   &0.004                  &-0.189                  &-0.298           
  &                        &                        &                        \\ 
       &$\phantom{\pm}\pm$0.039&$\phantom{\pm}\pm$0.033 &$\phantom{\pm}\pm$0.030
  &                        &                        &                 \\ \hline
  0.05 &                       &-0.095                  &
  &                        &                        &                        \\
       &                       &$\phantom{\pm}\pm$0.040 &              
  &                        &                        &                 \\ \hline
  0.10 &                       &-0.011                  &-0.129           
&                          &                        &                        \\
       &                       &$\phantom{\pm}\pm$0.048 &$\phantom{\pm}\pm$0.042
  &                        &                        &                 \\ \hline
  0.15 &                       & 0.057                  &-0.053           
&                          &                        &                        \\
       &                       &$\phantom{\pm}\pm$0.055 &$\phantom{\pm}\pm$0.049
  &                        &                        &                 \\ \hline
  0.20 &                       & 0.103                  & 0.019           
&-0.081                    &                        &                        \\ 
       &                       &$\phantom{\pm}\pm$0.058 &$\phantom{\pm}\pm$0.055
  &$\phantom{\pm}\pm$0.051 &                        &                 \\ \hline
  0.25 &                       &                        & 0.056           
&                          &                        &                        \\ 
       &                       &                        &$\phantom{\pm}\pm$0.051
  &                        &                        &                 \\ \hline
  0.30 &                       &                        & 0.084           
&0.0255                    &-0.051                  &                        \\ 
       &                       &                        &$\phantom{\pm}\pm$0.061
  &$\phantom{\pm}\pm$0.060 &$\phantom{\pm}\pm$0.057 &                 \\ \hline
  0.40 &                       &                        & 
& 0.060                    & 0.023                  &-0.020                  \\
       &                       &                        &              
  &$\phantom{\pm}\pm$0.063 &$\phantom{\pm}\pm$0.063 &$\phantom{\pm}\pm$0.061 
                                                                      \\ \hline
  0.50 &                       &                        &
&                        & 0.024                    & 0.024                  \\
       &                       &                        &              
  &                      &$\phantom{\pm}\pm$0.064   &$\phantom{\pm}\pm$0.064 
                                                                      \\ \hline
\hline
 \end{tabular}
 \end{center}
\caption{  Results of fits to \awf}
\label{tab-misas}
\end{table}
\end{small}

There is a definite pattern in the results:
\begin{itemize}
\item
 If the correct misassignment probability is assumed in analysis (the diagonal
  entries in the table), the result is unbiased, but the precision decreases
  rapidly with increasing $f_g$. For $f_g =0.5$, the result is identical with
  that obtained simply by folding production and both pairs of decay angles in
  the fit (an eightfold ambiguity), i.e. without applying any jet charge
  selection: $\awf =0.024 \pm 0.064$. For $f_g = 0.2$ (a value which might be
  expected in practice), the precision has already deteriorated to a value
  rather close to this.
\item
 If an incorrect misassignment probability is assumed in analysis, the result
  is biased. For instance, if the true misassignment rate is $f_g = 0.2$, but a
  value $f_a = 0.0$ is taken in analysis, a large negative bias in \awf\ is
  induced. This arises from the fact that the jet charge misassignment has had
  the effect of flattening the production angular distribution, which is also
  predicted for negative values of \awf. Unfortunately, in the region $f_g =
  0.2$, the bias seen in the table varies rather rapidly as the assumed
  misassignment probability diverges from its true value, implying that a large
  systematic error might be incurred in the fitted TGC value.
\end{itemize}

These considerations will have to be taken into account in determining the
usefulness of jet charge information in the analysis of the \Qqqq\ channel.

This above analysis assumes that the angle of the W is well measured.
We then wished to answer the question ``What effect will imperfections in
our definition of the W direction have on our ability to measure \awf\ ?''

The standard of way of estimating the W direction is to divide the tracks 
(and energy clusters)
of the event into 4 ``jets'', and to add the momenta of pairs of jets together 
to make the W momenta. The resulting directions are not aligned precisely 
with the directions of the Ws. 

In events where the reconstructed W is badly misaligned, 
the discrepancy can sometimes be traced to a misallocation
of the jets to the Ws. This leads naturally to the idea of a measurement 
degradation
due to a jet misassignment. However, this is an oversimplification. In
another class of badly measured events it is the ``jets'' themselves which
fail to measure the direction of the fundamental decay fermions. To be more
precise, in the context of \Pythia, if the sets of tracks which make up the
four jets is compared with the sets of particles which are the decay 
products of the four fermions, then for these events one or more of the
reconstructed jets will not be made up of particles from the decay of a
single fermion. 

The details of the problems caused by these effects may
depend rather sensitively on the apparatus used in the measurement and
the algorithm used to extract the W directions. For this reason 
we made no attempt
at the sort of generic measurement which is suitable
for the effects of jet charge. Instead, in order to provide an idea of 
the sort of degradation which follows from these considerations, we have taken
the example of the OPAL apparatus and processed EXCALIBUR events through a
a full simulation of the detector. The resulting simulated data is divided
into 4 jets using a Durham type algorithm; a kinematic fit is performed on
jets constraining the pairs to have the same mass, and the pairing with the
highest mass is taken as the correct one. The charge of each jet is defined
using a momentum weighted charge measure and the pair of jets which have the
highest summed jet charge is taken as the positive W.
For a dataset consisting of 7,000 fully simulated \Qqqq\ events we fit the 
value of \awf\ for all the events which are successfully reconstructed as 
four jet events. This defines a `realistic measure' of our ability to measure
\awf. 
Then the opening angles $\Omega_{j\phi}$ between each jet and fermion
are calculated, and the subset of events is taken where the minimum value
of $\Omega_{j\phi}$ allows a unique association between quark and jet. This
looses about 10\% of the data sample. For these events the W's direction is
calculated by taking the correct pairing as given by the generator information
and from that a value of \cw\ is derived. The charge assignment is then made by
the normal charge weighting methods. It turns out that for these events the
alignment between jet and quark is good, but the jet energies are not nearly 
so tightly constrained. Thus the W direction is much less well measured than
the individual quark directions. These are the ``good jet matching events''.
Finally we used the correct value of the W charge, and a value and error 
was calculated for these ``charge correct'' events. 

\begin{table}[htbp]
 \begin{center}
 \begin{tabular}{||c||c|c|c||}
\hline
 Data      & Reconstructed  & Pairing         & Charge         \\   
 Set       &                & correct         & correct        \\    
\hline
\hline
\awf\      &0.03-0.077+0.082&-0.18-0.072+0.080& 0.00$\pm$0.028 \\
\hline
 \end{tabular}
 \end{center}
\caption{  Results of fits to \awf}
\label{tab-fit}
\end{table}

The values of \awf\ which were found can be seen in Table \ref{tab-fit}, where
it can
be seen that the dominant cause of measurement degradation is the
mismeasurement of the W charge. Reanalysing the 90\% 
sample without any generator information produces a value fully consistent with
the original sample.

These values should not be extrapolated to other values of \awf. The accuracy
of the measurement of \awf\ depends on the value of \awf, and experimental
effects add towards that total error in ways which may also vary with \awf.

\section{CP violating TGC parameters}

In the usual phenomenological parametrization of the \tgc~\cite{LEP2YR}
the CP violating interactions are also included in the form: 
\bea
{\cal L}_{TGC}
&=&e\,g_{\scriptscriptstyle VWW} \Bigg [ 
g_4^V W^+_\nu W^-_\mu \left(\partial^\mu V^{\nu}+\partial^\nu V^{\mu}\right)
\nn\\
&+&i \tilde{\kappa}_V W^+_\nu W^-_\mu {\cal V}^{\mu\nu}
+i {\tilde{\lambda}_V\over m_W^2} 
{W^{+\mu}}_\nu W^-_{\rho\mu}{\cal V}^{\nu\rho} \Bigg ]
\label{lagrangian}
\eea
where
\[ V_{\mu\nu}=\partial_\mu V_{\nu}-\partial_\nu V_{\mu},\;\;
W^\pm_{\mu\nu}=\partial_\mu W^\pm_{\nu}-\partial_\nu W^\pm_{\mu},\;\;
\]
and 
\[ {\cal V}^{\mu\nu}=
\frac{1}{2}\varepsilon^{\mu\nu\rho\sigma} V_{\rho\sigma} .\;\;
\]
In Eq.(\ref{lagrangian}) $W^{\pm}$ is the $W$-boson field, and
the usual definitions $g_{\scriptscriptstyle\gamma WW}=1$, 
$g_{\scriptscriptstyle ZWW}=\mbox{ctg~}\theta_w$
are used. Finally  $e=g\sin\theta_w=g^\prime\cos\theta_w$.

What is the meaning of the above Lagrangian ? One way to answer this
question is to look at it as the low energy limit
of a manifestely $SU(2)_L\times U(1)_Y$ gauge invariant theory.
This can be done~\cite{HISZ,DERUJULA,GAUGEINVARIANT} 
in an effective Lagrangian approach
by considering gauge-invariant operators involving higher-dimensional
interactions among the gauge bosons and the Higgs field.
These operators
are scaled by an unknown parameter $\Lambda_{NP}$ describing 
the characteristic scale of some high energy New Physics,
generating at low energies the effective interaction 
$\L_{TGC}$ as a residual effect.
In order to generate all kinds of TGC appearing in Eq.(\ref{lagrangian}),
we need operators with dimension up to twelve.
On the other
hand, restricting ourselves to
$SU(2)_L\times U(1)_Y$-invariant operators with dimension six,
which are the lowest order ones in a $1/\Lambda_{NP}$ expansion, 
we end up with the following list of operators capable of inducing
CP-violating TGC couplings~\cite{GOUNARISPAPADOPOULOS}
\bea
\tilde{\OO}_{BW}&=&  \Phi^\dagger\;\frac{\vtau}{2} \cdot \vvW^{\mu\nu}
\Phi B_{\mu \nu}\nn\\
\tilde{\OO}_W&=&{1\over 3!}(\vW^{\mu}_{\;\rho}\times\vW^{\rho}_{\;\nu})
\cdot\vvW^{\nu}_{\;\mu} \ ,
\label{operator-cp}
\eea
where
\beq 
\tilde{B}^{\mu\nu}=\frac{1}{2}\varepsilon^{\mu\nu\rho\sigma}
B_{\rho\sigma}\;\;
\ , \ \;\;\vvW^{\mu\nu}=\frac{1}{2}\varepsilon^{\mu\nu\rho\sigma} 
\vW_{\rho\sigma}\;
\eeq
and $\Phi$ is the Higgs doublet.
\par
The New Physics contribution from the above operators is
described by the effective Lagrangian 
\beq
{\cal L}_{TGC}= \frac{g g^\prime}{2} {\dbw\over m_W^2}\OO_{BW} 
+ g{\dw\over m_W^2}\OO_W 
\label{gi-lagrangian}
\eeq
where the relations between $\dbw$, $\dw$,  
and the parameters appearing in the Eq.(\ref{lagrangian}) are given by
\bea
&\tk_\gamma = \dbw \;\;\;\;
\tl_\gamma =  \dw &           
\nn\\  &\tk_Z = -\tan^2\theta_w \tk_\gamma   \;\;\;\;
\tl_Z = \tl_\gamma& \ \ .
\label{relation}
\eea

In order to get an estimate of the expected sensitivity at LEP2, we performed
an analysis based on the event generator ERATO, using the method
of Optimal Observables, but without taking into account any
experimental effect such as resolution and related background. The results are
given in  Table \ref{cp-bound} corresponding to an 
integrated luminosity of 500 pb$^{-1}$.
\begin{table}[htb]
\begin{center}
\begin{tabular}{|c||c|c|c|c|}
\hline 
$\sqrt{s}$ (GeV) & 161 & 172 & 192 & 205  \\
 \hline\hline 
$\dbw$ & 1.81  & 0.79 & 0.08 & 0.08 \\
\hline 
$\dw$  & 0.43  & 0.14 & 0.02 & 0.016 \\
\hline
\end{tabular}
\caption[.]{One standard deviation errors on TGC parameters.}
\label{cp-bound}
\end{center}
\end{table}

It should be emphasised that the limits on the CP violating couplings
are expected to be of the same order as those on the CP conserving ones.

\section{High energy -vs- high luminosity}

The maximum centre-of-mass energy attainable at LEP is expected to be 
limited by the radio frequency accelerating gradient available.
As a result, a trade-off will be inevitable between the maximum
centre-of-mass energy achievable and the reliability of machine
operation, by keeping a fraction of the RF in reserve.
Other machine performance issues emphasise this point further.
There are therefore a range of strategies for running LEP during the
LEP2 programme.
The two extreme cases may be characterized as:
\begin{enumerate}
\item run at the highest energy attainable, with reduced safety factors, and
  accept the loss in integrated luminosity;
\item run at a lower energy and attempt to collect as much integrated
  luminosity as possible.
\end{enumerate}
In order to seek a balance between these extremes, one should
take into account the extra reach for searches and the expected precision of
measurements. 
One element in this accounting is the variation of the expected precision
for TGC measurements as a function of the centre-of-mass energy and the
integrated luminosity.

\subsection{Sensitivity studies with simulated events}

The prospective sensitivity to the anomalous couplings has been
studied for the case of the $\alpha$ parameters in the W$\phi$,
B$\phi$ and W models.
Two independent event generation and analysis procedures were applied
and consistent results were obtained.

In the first analysis, events were generated with the
EXCALIBUR~\cite{EXCALIBUR} program at various $\alpha$ values, and
analysed with a fitting program using the ERATO~\cite{ERATO} 
matrix element routines to calculate the differential cross-section.
The second analysis employed the WOPPER~\cite{WOPPER} event generator,
and analysed events using the on-shell cross-section 
formulae from Bilenky et al.~\cite{BILENKY}.
These two analysis strategies differ in sophistication, and 
include slightly different information
about the anomalous coupling sensitivity~\cite{LEP2YR}, 
but for the purpose of this study, i.e. the determination of the
sensitivity to the anomalous couplings, 
they give similar results.

Following~\cite{LEP2YR}, only the semileptonic channels 
$\mathrm{q}\mathrm{\overline{q}}\mathrm{e}\nu_{\mathrm{e}}$ and
$\mathrm{q}\mathrm{\overline{q}}\mu\nu_{\mu}$ were considered, and no
detector simulation was applied.
These ``ideal detector'' studies do not, of course, correspond
directly to real analyses of events observed in the LEP detectors,
because various additional effects enter: limited acceptances,
backgrounds, and detector resolution degrade the performance.
On the other hand, including information from the 
$\mathrm{q}\mathrm{\overline{q}}\tau\nu_{\tau}$,
$\mathrm{l}^+ \nu \mathrm{l}^- \bar{\nu}$ and
$\mathrm{q}\mathrm{\overline{q}}\mathrm{q}\mathrm{\overline{q}}$
channels improves the precision relative to the two-channel ideal
detector case.

It is possible to estimate the statistical precision that 
might be obtained in real analyses compared with these idealized 
studies by comparing the precision achieved and expected
from the small 172~GeV data sample already 
taken. The calibration factors were found to be within 20\% of unity within 
most cases, supporting the usefulness of the idealized studies. 
Only statistical errors were considered in these studies. While systematic 
effects may eventually become significant for these measurements, 
the dominant systematic contributions are expected to come from detector
effects, which can presumably be understood with sufficient precision with
sufficiently large data samples.

\begin{figure}
\begin{center}
\mbox{\epsfysize=12.0cm\epsffile{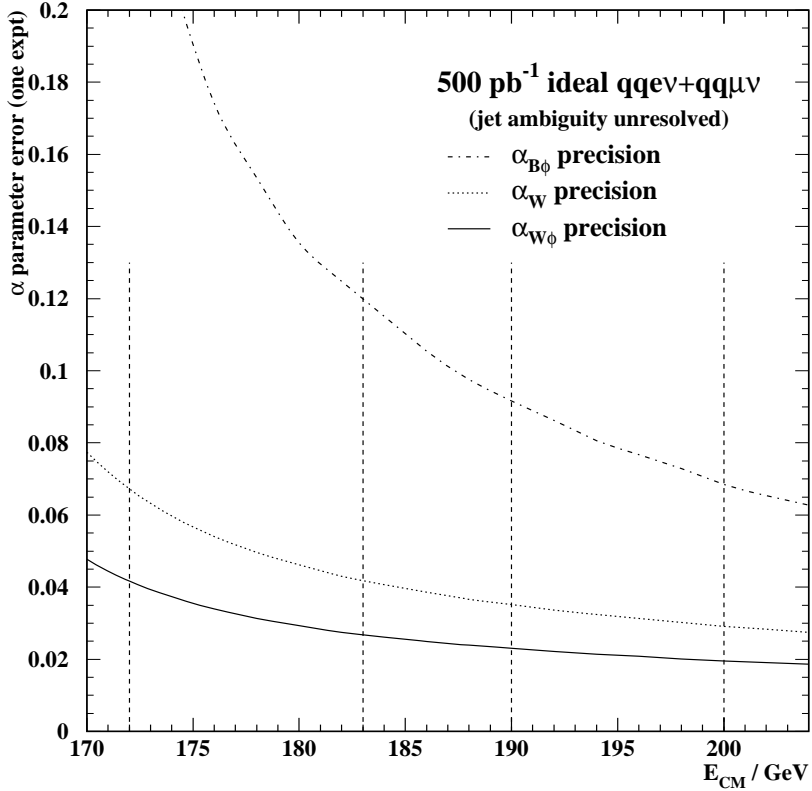}}
\vskip 0.5cm
\end{center}
\caption{
  Evolution of the expected statistical precision on the measurements of the
  anomalous coupling parameters $\alpha$ for different models, as a
  function of centre-of-mass energy. 
  The precisions given correspond to a single ideal detector analysis of
  the 
  $\mathrm{q}\mathrm{\overline{q}}\mathrm{e}\nu_{\mathrm{e}}$ and
  $\mathrm{q}\mathrm{\overline{q}}\mu\nu_{\mu}$ 
  channels, employing 
  the production and decay angles of both W's, but without resolving
  the two-fold decay ambiguity from the hadronically decaying W.
\label{fig:precision}
}
\end{figure}

Figure~\ref{fig:precision} shows the predicted error on the $\alpha$
parameters for centre-of-mass energies ranging from 170 GeV to 210
GeV, for 500 pb$^{-1}$ collected by a single experiment. 
The errors are derived for the case where the central value of the
$\alpha$ parameter is close to zero --- it was found that the expected
error decreases by ${\mathcal O}(30\%)$ if the true $\alpha$ parameter
is 1, for example, but the main effect is simply the substantially
increased W-pair production cross-section for such non-standard couplings.
The sensitivities at four canonical centre-of-mass energies are shown
in table~\ref{tab:future}. 
It is evident that the gain in precision with energy is very marked
below around 180~GeV, but that the gain then slows, although the
size of this effect differs for the different couplings.
It is interesting in particular to compare
the 190 and 200~GeV points: a unit of luminosity at 200~GeV is
approximately as sensitive as 1.4 to 1.8 units at 190~GeV.

\subsection{Conclusion}

The best measurement of anomalous coupling parameters at LEP2
requires that a large fraction of the luminosity should be taken above
180 GeV centre-of-mass energy, as is planned. 
It is important that the luminosity available should be high: the
500~pb$^{-1}$ considered in previous studies remains an apposite goal.
The sensitivity per unit luminosity rises above 180 GeV, so it is
useful to increase the centre-of-mass energy if only a modest cut in
luminosity is taken.
The sensitivity rise is, however, slow, so this analysis disfavours
operating strategies where a substantial luminosity penalty is
incurred for the sake of only a few GeV in centre-of-mass energy.

\begin{table}[p]
\begin{center}
\begin{tabular}{lccc} 
\hline
Centre-of-mass energy & $\Delta\alpha_{W\phi}$ & $\Delta\alpha_{B\phi}$ & $\Delta\alpha_{W}$ \\
\hline
172 GeV &  0.042 & 0.25  & 0.067 \\
183 GeV &  0.027 & 0.12  & 0.042 \\
190 GeV &  0.023 & 0.092 & 0.035 \\
200 GeV &  0.020 & 0.068 & 0.029 \\
\hline
\end{tabular} \\
\caption{Expectations for triple-gauge coupling precisions from
  a single experiment, approximated by the ``ideal'' precision
  for the 
  $\mathrm{q}\mathrm{\overline{q}}\mathrm{e}\nu_{\mathrm{e}}$ and
  $\mathrm{q}\mathrm{\overline{q}}\mu\nu_{\mu}$ channels.
  In each case the precision which would be obtained with 500
  pb$^{-1}$ at that centre-of-mass energy is given.}
\label{tab:future}
\end{center}
\end{table}

\section{Summary}

The first data which has been recorded at LEP2 above the W pair threshold 
has been followed by a rapid analysis by all of the LEP experiments to
search for possible signals for anomalous triple gauge boson couplings.
That this could be done so quickly is a tribute the the large amount
of preparatory work performed by theoreticians and experimentalists
in the provision of tools, and the development of analysis methods.
In this working group we reviewed our experience in the light of
this first data.
We primarily discussed the limitations which arise when trying to
incorporate all effects of W width, ISR, detector acceptance and 
resolution into an analysis. These effects had been previously shown to lead 
to biases, and several techniques have been developed to try to handle
this. Most methods do however have some limitations, and so we have considered
alternative ways to characterise the data which may be more amenable to the
use of analytic fitting tools. One of the most important outcomes
of the work was to nucleate the provision of a new fitting tool.
Work on this is still underway and will be published separately
at a later date.  
Another important topic is the extent to which use can be made of
W pair events decaying into the four quark channel. Analysis
of this channel is difficult due to the problems in identifying
the correct jet pairing and the correct W charge assignment, and
therefore a study was made to quantify the degradation 
in precision due to these effects.
We also briefly discussed the formulation of CP violating
\tgc\ parameters in a gauge invariant framework. Finally we
have presented a study of the trade off between achieving 
the highest possible LEP2 energy at the expense of integrated luminosity.


\end{document}